\documentclass[12pt]{article}

\usepackage{scicite}
\usepackage{color}
\usepackage{times}
\usepackage{graphicx}
\usepackage{longtable}
\usepackage{amssymb}
\newcommand{\apj}{Astrophys. J.}
\newcommand{\aj}{Astron. J.}
\newcommand{\apjs}{Astrophys. J., Suppl. Ser.}
\newcommand{\apjl}{Astrophys. J. Lett.}
\newcommand{\mnras}{Mon. Notices Royal Astron. Soc.}
\newcommand{\nat}{Nature}

\newcommand{\aap}{Astron. Astrophys.}
\newcommand{\aaps}{A\&{AS}}
\newcommand{\ssr}{Space Sci. Rev.}
\newcommand{\pasp}{Publ. Astron. Soc. Pac.}
\newcommand*{\arcsec}{^{\prime\prime}\mkern-1.2mu}
\usepackage{caption}
\usepackage{url}
\usepackage{longtable}

\newenvironment{sciabstract}{%
\begin{quote} \bf}
{\end{quote}}

\title{Optical polarization from colliding stellar stream shocks in a tidal disruption event}

\author
{I. Liodakis,$^{1\ast}$ K. I. I. Koljonen,$^{1,2,3}$ D. Blinov,$^{4,5}$  E. Lindfors,$^{1}$ K. D. Alexander,$^{6,7}$\\ T. Hovatta,$^{1,2}$ M. Berton,$^{1,2,10}$ A. Hajela,$^{6,8}$  J. Jormanainen,$^{1,9}$\\ K. Kouroumpatzakis,$^{4,5,11}$  N. Mandarakas,$^{4,5}$ K. Nilsson$^{1}$\\
\normalsize{$^{1}$Finnish Center for Astronomy with ESO, University of Turku, Turku, FI-20014, Finland}\\
\normalsize{$^{2}$Mets\"ahovi Radio Observatory, Aalto University, Kylm\"al\"a, FI-02540, Finland}\\
\normalsize{$^{3}$Institutt for Fysikk, Norwegian University of Science and Technology, Trondheim, 7491, Norway}\\
\normalsize{$^{4}$Institute of Astrophysics, FORTH - Hellas, Heraklion, GR-71110, Greece}\\
\normalsize{$^{5}$Department of Physics, University of Crete,  Heraklion, GR-70013, Greece}\\
\normalsize{$^{6}$CIERA, Northwestern University, Evanston, IL 60208, USA}\\
\normalsize{$^{7}$Steward Observatory, University of Arizona, Tucson, AZ 85721-0065, USA}\\
\normalsize{$^{8}$DARK, Niels Bohr Institute, University of Copenhagen, 2200 Copenhagen, Denmark}\\
\normalsize{$^{9}$Department of Physics and Astronomy, University of Turku, Turku, FI-20014, Finland }\\
\normalsize{$^{10}$European Southern Observatory, Santiago, 19001, Chile}\\
\normalsize{$^{11}$Astronomical Institute, Academy of Sciences, Prague, CZ-14131, Czech Republic}\\
\\
\normalsize{$^\ast$E-mail: yannis.liodakis@utu.fi.}}

\date{}

\begin{document} 


\baselineskip24pt


\maketitle


\begin{sciabstract}
A tidal disruption event (TDE) occurs when a supermassive black hole rips apart a passing star. A part of the stellar matter falls toward the black hole, eventually forming an accretion disk that occasionally launches a relativistic jet. We performed optical polarimetry observations of a TDE, AT 2020mot, finding a peak linear polarization degree of 25$\pm$4\% indicative of highly polarized synchrotron radiation typically observed from relativistic jet sources. However, we do not find a comparable radio emission expected from such systems even eight months after the optical peak. Instead, we suggest that the linearly polarized optical emission arises from shocks that form in the accretion disk formation process as the stellar matter stream collides with itself.

\end{sciabstract}

A tidal disruption event occurs when a star passes close enough to a supermassive black hole (BH) for the star's self-gravity to be overpowered by tidal forces produced by the BH's gravitational potential \cite{Rees1988}. As the star is torn apart, roughly half of the gas forms an accretion disk before falling into the BH. The disruption leads to an outburst of emission which peaks in the ultraviolet (UV) and soft X-rays, with a spectrum corresponding to a $\sim10^5$ K black body. This is followed by a steady decline, regulated by the rate gas fall on the black hole, with temporal (t) evolution  proportional to $\sim{t}^{-5/3}$ \cite{Rees1988}. Multiwavelength observations of TDEs have shown that some lead to the BH launching a relativistic jet of plasma \cite{Alexander2020}. TDEs can therefore provide information on the early stages of formation for both accretion disks and jets. 

Some TDEs, known as optical TDEs \cite{VanVelzen2020}, have emission that peaks in the optical-UV range, due to having a lower black body temperature than typical TDEs. Optical TDEs are typically faint or undetected in the radio and X-rays and do not show evidence of a jet. The emission profile of such events is not consistent with the standard theoretical model of TDEs [also known as X-ray TDEs, \cite{Rees1988}]; several alternative models have been proposed to explain this divergence. Possible explanations include X-ray emission reprocessed to UV-optical wavelengths by surrounding gas \cite{Metzger2016} or emission powered by tidal shocks that form due to the star's gas stream colliding with itself as it flows around the black hole  during accretion disk formation [known as the outer shock model, \cite{Shiokawa2015}]. There is observational evidence in favor of both models. For example, the brightness evolution and spectrum of the TDE AT 2018fyk  have been interpreted as evidence for reprocessed emission \cite{Wevers2019}. Similar conclusions have been reached for two other TDEs, AT 2018hyz and AT 2019qiz, based on their spectral line properties \cite{Hung2020,Nicholl2020}. In contrast, the late-time X-ray brightening of the TDEs ASASSN-15oi and AT 2019azh, have been interpreted as due to the late accretion disk formation expected in the outer shock model \cite{Gezari2017,Liu2019}.

Reprocessing and outer shock models make similar predictions for the total emission intensity, light curve, and late-time emission (e.g., X-ray brightening). However, their polarization properties are expected to differ substantially. Previous polarization studies have reported a single detection of $\sim7-8\%$ polarized near-peak emission for TDEs with jets and $\sim1-2\%$ for optical TDEs \cite{Wiersema2012,Higgins2019,Lee2020,Wiersema2020}. Other observations have yielded only upper limits. Polarization measurements could potentially be used to determine the orientation of the accretion disk or the jet magnetic field.

\subsection*{Observations of AT 2020mot}

We measured the polarization of AT 2020mot (also known as ZTF20abfcszi or Gaia20ead), an optical TDE identified in June 2020 in a nearby galaxy at a redshift $z=0.07$ \cite{ztfsurvey2019}. It was classified as a TDE because it showed broad He~\textsc{ii} and hydrogen (Balmer) emission lines in the optical spectrum \cite{2020TNSCR2478....1H}. We supplemented our polarization observations with radio, X-ray, photometric, and spectroscopic observations up to eight months from the optical peak  \cite{Methods}. Our radio observations with the Karl G. Jansky Very Large Array (VLA) and public X-ray observations with the  Neil Gehrels {\it Swift} Observatory ({\it Swift})  did not detect the transient source, which is not uncommon for an optical TDE. Near the optical peak (Modified Julian Date [MJD] 59077.8) the radio 3$\sigma$ upper limit was $<27$ micro-Jansky ($\rm \mu\mathrm{Jy}$) at 15~GHz. Radio observations at lower frequencies (1-7~GHz) more than six months after the optical peak detect a point source showing an optically thin power-law spectrum, that is consistent with the non-detection at 15~GHz, and indicates synchrotron emission from star formation. The star-formation rate (SFR) calculated from the radio observations differs by 2.3$\sigma$ from the star-formation rate estimated using infrared data. During the decay phase, an X-ray source was detected with flux of 0.0016$\pm$0.0004 $\rm counts\cdot s^{-1}$ from the direction of AT 2020mot. Our observations with {\it Swift} and {\it XMM}-Newton six to eight months (MJD 59270.5, 59306.4, and 59308.4) after the peak did not find any rise in the X-ray emission. Instead, we attribute all the X-ray emission to a contaminating source, likely a background active galactic nuclei (AGN). The measured UV and optical fluxes six to eight months post-peak do not indicate any elevated ultraviolet activity above the background level of the host galaxy level. Additional spectroscopic observations using the Nordic Optical Telescope (NOT) did not show any optical emission lines associated with the TDE about six months after the outburst peak (MJD~59264). All these observed properties are consistent with AT 2020mot being a typical optically-selected TDE \cite{Methods}.

We began polarization observations of AT 2020mot approximately 25 days after the outburst peak, using the RoboPol polarimeter mounted on the 1.3m telescope of the Skinakas Observatory, Crete \cite{Ram2019}. We used RoboPol to observe the source three times separated by roughly 20 days. We performed a fourth polarization observation using NOT before the $R$-band optical flux reached the host galaxy level (MJD 59164). Figure \ref{fig:lightcurve} shows the ZTF $r$-band light curve \cite{Smith_2019}, the evolution of the polarization degree ($\Pi$) and the polarization angle (PA). We measured a low polarization degree ($\Pi\sim$ 2\%) for the first observation, similar to the values previously reported from other optical TDEs \cite{Higgins2019}. The second observation showed a higher level of polarization ($\Pi\sim7$\%) similar to TDEs with a jet, accompanied by a change in the PA. The last two observations showed a decline in $\Pi$ with a stable PA (within the uncertainties). After removing the unpolarized contribution of the host galaxy, we calculate that the source reached an intrinsic $\Pi=25\pm4\%$ at the time of the second observation.

\subsection*{Interpretation of the optical polarization}

The high polarization degree observed in AT 2020mot is most likely to be produced by synchrotron radiation. We consider whether this could be due to the formation of a relativistic jet, which could explain both the observed polarization level and its variability. The lack of both radio and X-ray detections near the peak of the TDE allows us to reject a collimated jet viewed at a small angle to the line of sight. This is consistent with the rarity of TDEs with jets \cite{Alexander2020}. On-axis gamma-ray burst jets, which are the only other transient short-lived jet source, have been observed to have high $\Pi$ near the $\gamma$-ray peak, which steadily declines over time with either a stable PA or exhibiting 90$^o$ jumps  \cite{Sari1999,Mundell2013,Wiersema2014}, which we do not observe. It is possible that an off-axis jet could produce radio and X-ray emission that is delayed with respect to the outburst peak \cite{Mattila2018}. However, we expect the jet opening angle would increase over time \cite{Piran2004}, causing a late-time brightening in both  radio and X-ray emission. There was no such increase in brightness at the time of our observations. An off-axis jet would not contribute substantially to the emission at early times, so could not produce the observed $\Pi=25\%$ even if the jet was maximally polarized.

Scattering from an accretion disk can polarize optical emission. The maximum degree of polarization that can be produced by scattering is $\Pi=11.7\%$ for an exactly edge-on disk, and rapidly declines towards a pole-on oriented disk \cite{Chandrasekhar1960}. In a disk scattering scenario, we would expect the PA to remain at a constant value, set by the scattering geometry. Both the higher value of $\Pi$ and the PA variability lead us to  reject accretion disk scattering as the origin of the polarized emission.

In an alternative reprocessing scenario, polarized emission would arise from radiation emitted by an accretion disk that is scattered by the optically thick obscuring material left behind from the disrupted star. In this case, a clumpy medium could cause polarization angle variability, but the polarization degree is expected to be $<10\%$ \cite{Marin2015}. Reprocessing models often invoke an accretion disk wind for the obscuring material. The quasi-spherical geometry of the disk wind is also expected have low polarization, due to symmetry. The disk wind is also expected to produce detectable radio emission \cite{Matsumoto2021} which we do not find. Nor was there any UV or X-ray enhancement or spectral lines associated with reprocessing. We therefore reject the reprocessing scenario as the origin of the optical emission from AT 2020mot.

Another possibility is stellar stream shocks \cite{Piran2015,Shiokawa2015}. In this scenario, the gas from the disrupted star does not immediately form an accretion disk, as in the standard TDE model \cite{Rees1988}. Instead, the stellar stream flowing around the BH collides with itself. The collision forms shocks that help in dissipating the energy of the flow of stellar material and circularizes it towards the formation of the accretion disk, and in the process account for all of the optical emission.

\subsection*{Colliding stellar streams model}  
  
Figure \ref{fig:cartoon} shows a schematic of our proposed scenario for AT 2020mot. The UV and optical spectral energy distributions of optical TDEs are usually consistent with a nearly constant temperature black body model \cite{Hung17}. However, we found that in a few different times a power-law model provides a better fit to the spectral energy distributions of AT 2020mot. This indicates that the UV-optical emission comes from either a modified black body or a combination of several emission components \cite{Methods}. To obtain an estimate for the bolometric luminosity of AT 2020mot at the peak of the outburst, we fitted the UV-optical spectrum of the peak with a black body model, finding a luminosity of $(1\pm0.2)\times10^{44}$ erg$\rm\cdot s^{-1}$ and a temperature of $21,800\pm1600$ K.

\begin{figure}
\centering
\includegraphics[scale=0.4]{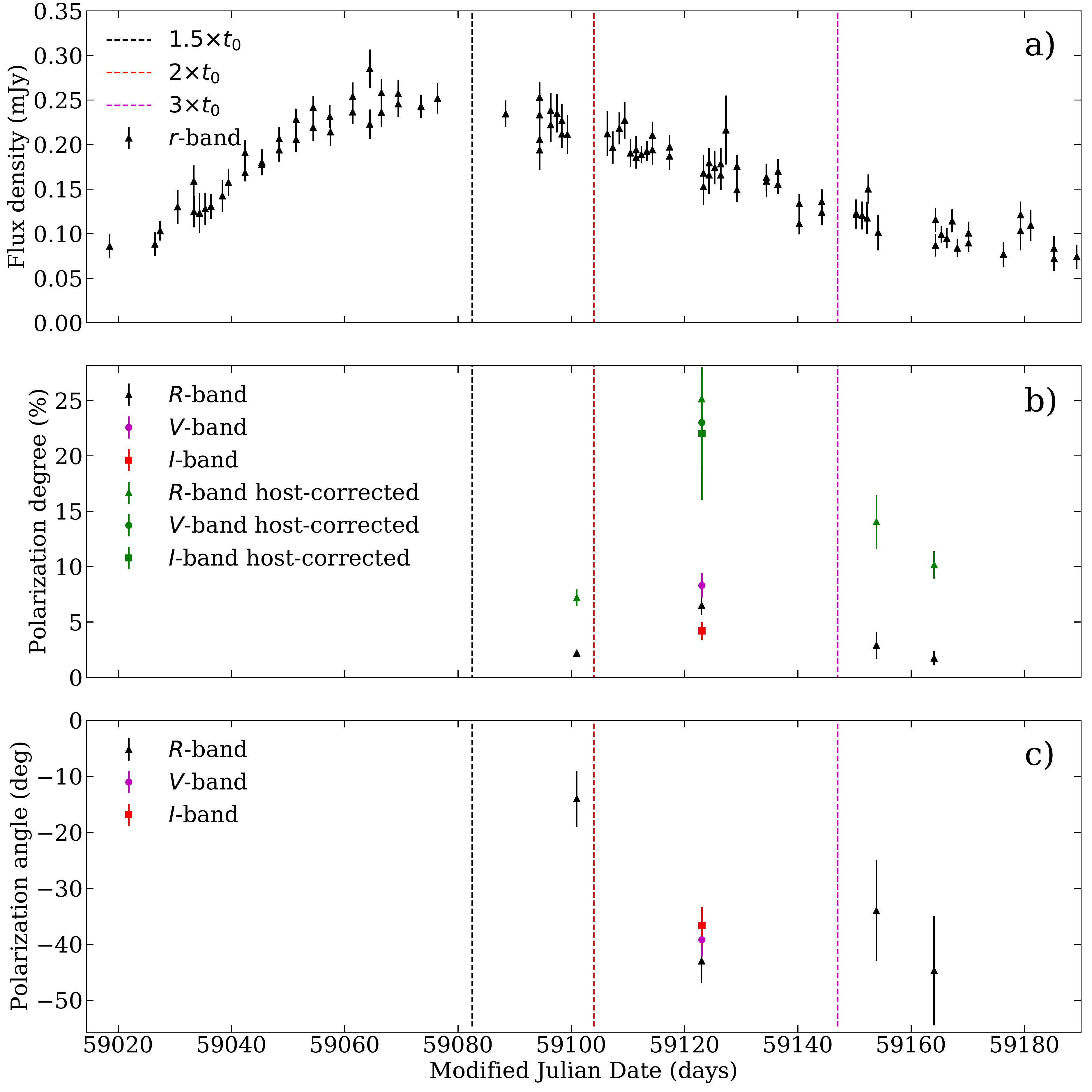}
\caption{{\bf Optical light curve and polarization measurements of AT 2020mot.} (a) $r$-band brightness (black triangles), showing the temporal evolution. (b) Observed $V$-band (purple circles), $R$-band (black triangles), and $I$-band (red squares) and host-galaxy-corrected polarization degree for the $V$-band (green circle), $R$-band (green triangles), and $I$-band (green square). (c) polarization angle, with the same symbols. The errorbars indicate the 68\% confidence interval. In all panels, the vertical dashed lines show the characteristic times calculated using the outer shock model for $1.5\times t_0$ (black), $2\times t_0$ (red), and $3\times t_0$ (magenta). We infer that the disruption happened before the rise of the $r$-band light curve, at $t<59026$ MJD.}
\label{fig:lightcurve}
\end{figure}

\begin{figure}
\centering
\includegraphics[scale=0.7]{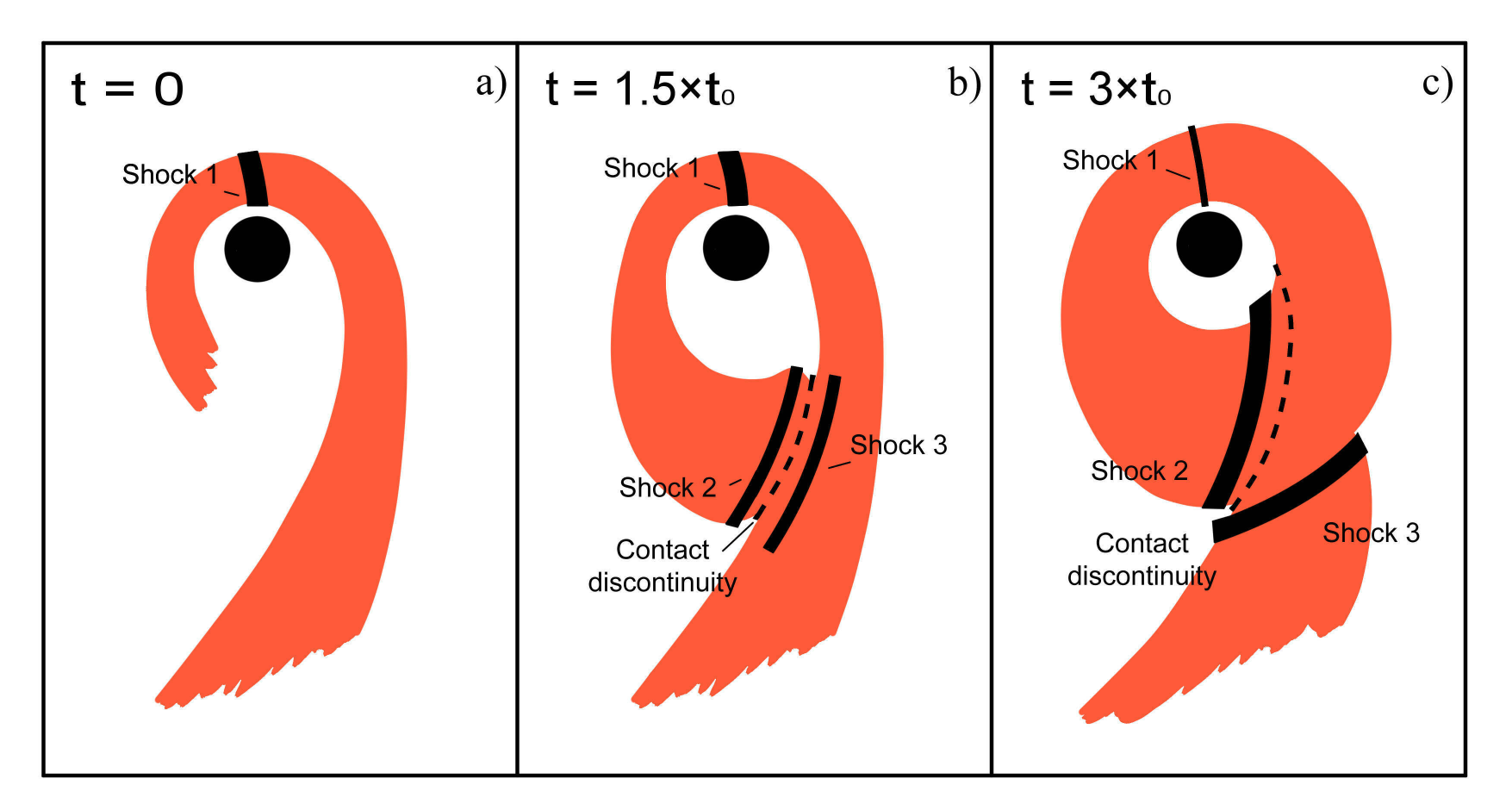}
\caption{{\bf Schematic qualitative diagram of the tidal stream shock model [after  \cite{Piran2015}]}. Each panel shows a different time. (A) Shock 1 forms after the stellar disruption and dominates the optical emission during the rise of the light curve. (B) By 1.5$\times{t_0}$, two additional shocks have formed, and the light curve reaches its peak brightness (Fig. \ref{fig:lightcurve}a). As time passes, shocks 2 \& 3 account for increasing fractions of the emission. (C) By 3$\times{t_0}$ shocks 2 \& 3 dominate the total emission. Between 1.5$\times{t_0}$ and 3$\times{t_0}$, the changing shock orientation and relative contributions to the optical emission cause the polarization degree and polarization angle to vary.}
\label{fig:cartoon}
\end{figure}

Applying these values to a previously-published outer shock model \cite{Ryu2020}, we estimate a BH mass of $3.6\times{10^6}$ solar masses ($ M_\odot$), disrupted star mass $M_\star=1.3~M_\odot$, and orbital timescale of the inner most material $t_0\approx43$ days. Figure \ref{fig:lightcurve} indicates the predicted peak position at 1.5$\times{t_0}$, which is consistent with the observed light curve. In the outer shock model, the emission during the initial rise is dominated by the shock formed at the pericenter (also known as the nozzle shock, Fig. \ref{fig:cartoon}a) of the stellar stream orbit [\cite{Piran2015}, labelled shock 1]. At the outburst peak (Fig. \ref{fig:cartoon}b), the reverse and forward shocks [the outer shocks, labeled shocks 2 \& 3] have  formed. By 2$\times{t_0}$  (Fig. \ref{fig:lightcurve}), the outer shocks account for a larger fraction of the optical emission, which coincides with when we measure the highest $\Pi$ value and the change in PA. The degree of polarization produced by each shock depends on the shock compression and its orientation with respect to the line of sight \cite{Methods}. The observed polarization behavior can then be produced as the sum of multiple shocks with differing PAs and varying contributions to the total emission. Changes in orientation of the shocks, as well as changes in their relative contributions, could produce large variations in $\Pi$, similar to the behavior observed due to presence of multiple shocks in AGN jets \cite{Liodakis2020}. By 3$\times{t_0}$  (Fig. \ref{fig:cartoon}c), the emission is dominated by the outer shocks. Subsequent weakening of the shocks, an increase in turbulence, or the break-up of shock 2 could then produce the observed decrease in the observed $\Pi$. The evolution of the optical-to-UV spectrum \cite{Methods} is also consistent with a varying contribution to the emission from different shocks.

We strongly favor tidal stream shocks as the origin of the optical emission in AT 2020mot. In this scenario, the lack of late time UV or X-ray brightening can be interpreted as demonstrating that the system did not form an accretion disk even eight months after the optical peak. However, a weak accretion disk emitting at a lower flux than the host galaxy cannot be excluded by the observations. AT 2020mot is similar to other optical TDEs, so we suggest that the tidal stream shock model might explain the difference between optical and X-ray TDEs.

\newpage





\section*{Acknowledgments}

We thank  N. Globus, N. Kylafis, T. Piran, and R. Romani for comments and suggestions that helped improve this work. I.L. thanks the University of Crete and Institute of Astrophysics for their hospitality while this paper was written. Based on observations obtained with XMM-Newton, an ESA science mission with instruments and contributions directly funded by ESA Member States and NASA. Lasair is supported by the UKRI Science and Technology Facilities Council and is a collaboration between the University of Edinburgh (grant ST/N002512/1) and Queen’s University Belfast (grant ST/N002520/1) within the LSST:UK Science Consortium. ZTF is supported by National Science Foundation grant AST-1440341 and a collaboration including Caltech, IPAC, the Weizmann Institute for Science, the Oskar Klein Center at Stockholm University, the University of Maryland, the University of Washington, Deutsches Elektronen-Synchrotron and Humboldt University, Los Alamos National Laboratories, the TANGO Consortium of Taiwan, the University of Wisconsin at Milwaukee, and Lawrence Berkeley National Laboratories. Operations are conducted by COO, IPAC, and UW. This research has made use of the Aladin sky atlas developed at CDS, Strasbourg Observatory, France \cite{Bonnarel2000} and \cite{2014ASPC..485..277B}. We acknowledge the use of public data from the {\it Swift} data archive.
Based on observations made with the Nordic Optical Telescope, owned in collaboration by the University of Turku and Aarhus University, and operated jointly by Aarhus University, the University of Turku and the University of Oslo, representing Denmark, Finland and Norway, the University of Iceland and Stockholm University at the Observatorio del Roque de los Muchachos, La Palma, Spain, of the Instituto de Astrofisica de Canarias. Part of the Nordic Optical Telescope observations were performed by the 2020 Nordic Optical Telescope observing school for upper secondary school students, organised by the Science Centre of the University of Turku and the Finnish Centre for Astronomy with ESO, University of Turku.
This research has made use of data and/or software provided by the High Energy Astrophysics Science Archive Research Center (HEASARC), which is a service of the Astrophysics Science Division at NASA/GSFC. The Pan-STARRS1 Surveys (PS1) and the PS1 public science archive have been made possible through contributions by the Institute for Astronomy, the University of Hawaii, the Pan-STARRS Project Office, the Max-Planck Society and its participating institutes, the Max Planck Institute for Astronomy, Heidelberg and the Max Planck Institute for Extraterrestrial Physics, Garching, The Johns Hopkins University, Durham University, the University of Edinburgh, the Queen's University Belfast, the Harvard-Smithsonian Center for Astrophysics, the Las Cumbres Observatory Global Telescope Network Incorporated, the National Central University of Taiwan, the Space Telescope Science Institute, the National Aeronautics and Space Administration under Grant No. NNX08AR22G issued through the Planetary Science Division of the NASA Science Mission Directorate, the National Science Foundation Grant No. AST-1238877, the University of Maryland, Eotvos Lorand University (ELTE), the Los Alamos National Laboratory, and the Gordon and Betty Moore Foundation. This research has made use of data and/or software provided by the High Energy Astrophysics Science Archive Research Center (HEASARC), which is a service of the Astrophysics Science Division at NASA/GSFC. This project has received funding from the European Research Council (ERC) under the European Union’s Horizon 2020 research and innovation programme (grant agreement No. 101002352). This publication makes use of data products from the Two Micron All  Sky Survey, which is a joint project of the University of Massachusetts and the Infrared Processing and Analysis Center/California Institute of Technology, funded by the National Aeronautics and Space Administration and the National Science Foundation. We acknowledge funding to support our NOT observations from the Finnish Centre for Astronomy with ESO (FINCA), University of Turku, Finland (Academy of Finland grant nr 306531).

\subsection*{Funding}
EL was supported by Academy of Finland projects 317636 and 320045.
KIIK was supported by Academy of Finland project 320045 and the European Research Council (ERC) under the European Union’s Horizon 2020 research and innovation programme (grant agreement No. 101002352).
TH was supported by Academy of Finland projects 317383, 320085, and 322535.
JJ was supported by Academy of Finland project 320085.
DB and NM acknowledge funding from the European Research Council (ERC) under the European Unions Horizon 2020 research and innovation programme under grant agreement No. 771282. 
KDA was supported by NASA through the NASA Hubble Fellowship grant \#HST-HF2-51403.001-A awarded by the Space Telescope Science Institute, which is operated by the Association of Universities for Research in Astronomy, Inc., for NASA, under contract NAS5-26555.

\subsection*{Author contributions}
DB, TH, JJ, EL, IL, and NM performed the polarization observations, data analysis and modeling. KIIK performed the UV and X-ray analysis. KDA and AH performed the radio observations and data reduction. MB, EL, K. Kour. and KN performed the host-galaxy characterization, optical spectroscopy, and star-formation rate estimation. K. Kour performed the spectral energy distribution fitting. IL, KIIK and KN estimated the black hole mass. All authors contributed writing the paper and the interpretation of the results.

\subsection*{Competing interests}
The authors declare that they have no competing interests.

\subsection*{Data and materials availability}

Our RoboPol and NOT observations are available on Zenodo  \cite{RoboPol_data,NOT_data}. Our radio observations are available at \url{https://data.nrao.edu/} under project code '20A-372'. The archival optical photometry is available at \url{https://lasair-ztf.lsst.ac.uk/objects/ZTF20abfcszi/}, and the archival UV \& X-ray data at \url{https://heasarc.gsfc.nasa.gov/cgi-bin/W3Browse/w3browse.pl} by searching for object name 'AT 2020mot' and setting the mission to 'Swift' and 'XMM-Newton'.


\newpage

\makeatletter 
\renewcommand{\thefigure}{S\@arabic\c@figure}
\renewcommand{\thetable}{S\@arabic\c@table}
\renewcommand{\thepage}{S\@arabic\c@page}
\makeatother

\setcounter{figure}{0}

\setcounter{table}{0}

\setcounter{page}{0}

\thispagestyle{empty} 

\begin{center}
{\Large Supplementary Materials for}\\
\bigskip
{ \bf Optical polarization from colliding stellar stream shocks in a tidal disruption event}\\
\bigskip
{I. Liodakis, K. I. I. Koljonen, D. Blinov,  E. Lindfors, K. D. Alexander, T. Hovatta, M. Berton, A. Hajela,  J. Jormanainen, K. Kouroumpatzakis,  N. Mandarakas, K. Nilsson}\\

Corresponding author: yannis.liodakis@utu.fi
\end{center}

\bigskip
\bigskip
\bigskip

\noindent {\bf The PDF file includes:}\\
\noindent Materials and Methods\\
\noindent Figs. S1 to S8\\
\noindent Tables S1 to S4\\
\noindent References \textit{(36-82)}



\newpage

\subsection*{Materials and Methods}

\subsubsection*{Optical polarization}

Optical polarization measurements of the first three epochs (2020 Sep 8; 2020 Oct 1 and 31) were obtained with the RoboPol polarimeter at the 1.3-m telescope of the Skinakas observatory \cite{Ram2019}. The measurements were processed with a fixed $4.5\arcsec$ aperture using the standard pipeline \cite{King2014,Panopoulou2015} with modifications described elsewhere \cite{Blinov2021}. The following set of polarization standards was observed multiple times during the same nights as AT 2020mot as well as during adjacent nights: BD +32 3739, HD 154892, BD +28 4211, BD +59 389, HD 212311, BD +64 106, BD +33 2642. Measurements of these standards provided corrections of the instrumental polarization and the zero point of the polarization position angle. Their uncertainties were propagated to the final AT 2020mot polarization parameters estimates. The fourth polarization observation was performed with the Alhambra Faint Object Spectrograph and Camera (ALFOSC) at the Nordic Optical Telescope on 2020 November 10. The observations were carried out in the $R$-band using the standard setup for linear polarization observations, i.e., a half-wave plate followed by a birefringent crystal, in this case a calcite block. The observations were performed in non-optimal weather conditions (thin clouds). For the data calibration, we observed two highly-polarized standards stars (HD 19820, BD +25 727) taken immediately after the observation of the source.

The data were analysed using the semi-automatic pipeline developed at the Tuorla Observatory \cite{Nilsson2018}. This pipeline uses standard procedures: the sky-subtracted target counts were measured in the ordinary and extraordinary beams using aperture photometry. The normalized Stokes parameters and the polarization fraction and position angle were calculated from the intensity ratios of the two beams using standard formula \cite{Landi2007}. For consistency, we used the same aperture size as the RoboPol observations ($4.5\arcsec$). Table \ref{tab:opt_pol} lists the polarization and Stokes parameters from our observing campaign.

\bigskip

\begin{table}
\caption{{\bf Optical linear polarization observations of AT~2020mot}. The first and second columns show the date of the observation an the optical band used. The third and fourth columns the polarization degree and angle,and the last two columns the normalized Stokes q and Stokes u parameters.}
\label{tab:opt_pol}
\begin{center}
\begin{tabular}{lccccc}
        \hline
         Date & Band & $\Pi$ & PA & q & u \\
         (MJD) &  &  (\%) & (degrees) & &\\
         \hline 
         59101.45321 & $R$ & 2.2$\pm$0.3 &  -14$\pm$5 &         0.019 $\pm$ 0.003 &  -0.010 $\pm$ 0.004 \\
         \hline
         59123.53090 & $V$ &   8.3 $\pm$ 1.1    &   -39.2  $\pm$  3.0    &  0.017 $\pm$ 0.009 &  -0.081 $\pm$ 0.011\\
         59123.50614 & $R$ &  6.5 $\pm$ 0.9    &   -43  $\pm$  4    &     0.005 $\pm$ 0.009 &  -0.065 $\pm$ 0.009\\
         59123.55194 & $I$ &  4.2 $\pm$ 0.8   &    -36.7 $\pm$3.4  &     0.012 $\pm$ 0.005  & -0.041 $\pm$  0.008\\
         \hline
         59154.35861  & $R$ &  2.9  $\pm$  1.2    &   -34  $\pm$   9    &    0.010 $\pm$   0.008  &  -0.027 $\pm$  0.013\\
          \hline
         59164.55017  & $R$ &  1.75 $\pm$ 0.64   & -44.7 $\pm$  9.8    &    0.000 $\pm$  0.006   &   -0.018  $\pm$  0.006\\

         \hline\hline
\end{tabular}
\end{center}
\end{table}

\bigskip
The unpolarized flux of the host galaxy within the aperture dilutes the polarization signal such that the measured polarization is always lower than the intrinsic polarization \cite{2008MNRAS.388.1766A}. To correct for this effect, we must estimate the host galaxy flux, $I_\mathrm{host}$, its effective radius (to determine how much of this flux is within the aperture) and the total intensity, $I$. We derive the host galaxy fluxes in different bands from the Panoramic Survey Telescope and Rapid Response System (Pan-STARRS) pre-TDE magnitudes \cite{PANSTARRS_database}, using filter transformations from \cite{2012ApJ...750...99T}. We determine the effective radius of the host galaxy from our imaging observations with the NOT performed at the same time as the polarimetric (see below). Because the galaxy is very compact, its $R$-band flux of $I_\mathrm{host,R}=0.48\pm0.05$ \,mJy is fully contained within our aperture of $4.5\arcsec$. We estimate the intrinsic, host-galaxy corrected,  polarization degree ($\Pi_\mathrm{corr}$) following \cite{Hovatta2016} as $\Pi_\mathrm{corr}= \Pi_\mathrm{obs}\times{I}/(I-I_\mathrm{host})$, where $\Pi_\mathrm{obs}$ is the observed polarization degree. For comparison, we measured $I$ in three different ways: i) direct RoboPol photometry extracted from the polarimetric observations; ii) $V$-band photometry from UVOT (see below); iii) using the nearest Zwicky Transient Facility (ZTF) $r$-band flux density observation from LASAIR (within 1 day) to the polarization observations. In the latter case $I$ is estimated as $I=I_\mathrm{TDE}+ I_\mathrm{host}$. Those yield $\Pi_R=22\pm5\%$, $\Pi_V=23\pm4\%$, and $\Pi_R=25\pm4\%$ respectively. Therefore, all our $\Pi_\mathrm{corr}$ measurements agree within 1$\sigma$. The uncertainty $\sigma_{\Pi_\mathrm{corr}}$ is estimated by accounting for the uncertainty in the total flux ($\sigma_I$), uncertainty in the host galaxy flux ($\sigma_{I_\mathrm{host}}$), and the observed polarization degree ($\sigma_\Pi$). This is achieved by $10^3$ random samplings from Gaussian distributions with the mean corresponding to the estimated value and the standard deviation to the corresponding individual $\sigma$. For the night of 2020 October 1 (MJD~59123.05), we also corrected for the host-galaxy contribution in the $I$-band. We find $\Pi_I=22\pm6\%$ consistent with the $V$-band and $R$-band values.

The small residuals of the elliptical profile model that we fitted to the host-galaxy (see below) indicates that an edge-on orientation is unlikely. This implies that any dust scattering induced polarisation from the host galaxy is $<1\%$ \cite{Simmons2000,Jones2012}. We estimate the reddening to the line of sight  AT~2020mot to be $E(B-V)=0.088$ \cite{schlafly11}.  Using standard relations \cite{Hiltner1956,Serkowski1975} we estimate an upper limit on the interstellar polarization of $\Pi<0.63\%$. In an alternative scenario \cite{Panopoulou2019} the maximum possible interstellar polarization is $\Pi<0.91\%$. In both cases the expected level of induced polarization is much smaller than  AT 2020mot's $\Pi$ measurements, and smaller than our measurement uncertainty, hence our observed $\Pi$ is intrinsic to the TDE.

The origin of the polarized emission is uncertain given the many unknowns in the system. The maximum observed polarization of 25\% is hard to reconcile with the majority of known mechanisms, and it is most likely synchrotron emission. This expected to be emitted by TDEs because the star's magnetic field is amplified at the shock front \cite{Bonnerot2017}. Synchrotron emission from a thermal distribution of electrons has been shown to produce high degree of polarization \cite{Pandya2016}. For a standing shock the degree of polarization will depend on the shock compression and the viewing angle ($\theta$) of the shock towards the observer. For $\theta>45^o$ and a low Mach number ($\mathcal{M}$), the polarization degree can be $>25\%$ for a perfectly ordered magnetic field. We expect shock 3 to be a strong shock [$\mathcal{M}\sim10$, \cite{Shiokawa2015}], so would produce similar or  higher level of polarization.

\subsubsection*{Radio observations}

\begin{figure}
\centering
\includegraphics[scale=3]{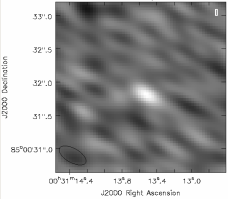}
\caption{{\bf Radio image of AT~2020mot with the VLA.} The observation was taken in the $C$-band (4-8~GHz) on the 28 February 2021. The black oval in the lower left corner of the image shows the size and shape of the telescope beam. AT~2020mot is in the center of the image. A point source model for the flux density is preferred to an extended Gaussian model for the source at all frequencies.} 
\label{fig:radio}
\end{figure}

We observed AT 2020mot with the NSF's Karl G. Jansky Very Large Array (VLA) on 2020 Aug 16, 2021 Jan 29, and 2021 Feb 28. We reduced the data in \textsc{CASA} \cite{casa} using standard procedures and imaged the data using \texttt{CLEAN}. In our first observation, we do not detect any emission at the TDE position, with a $3\sigma$ upper limit of $<27~\mu$Jy at a mean frequency of 15 GHz. In our second observation, we detect an unresolved source with a flux density of $60\pm8$ $\mu$Jy at a mean frequency of 6 GHz (Fig. \ref{fig:radio}); we then split the data into two frequency bins and reimage estimate the spectral index (Table \ref{tab:radio_obs}). After this detection, we observed a final epoch spanning 1 to 8~GHz, to better constrain the spectral index of the emission and check for variability. The source is detected at all frequencies and we split the data by frequency into 1 to 2~GHz bandwidth chunks for imaging. To measure the flux density and uncertainty, we fitted a point source model in the image plane using the {\tt fitsrc} command within the {\tt imtool} package of {\tt pwkit} \cite{pwkit}. The measured flux density at 1.4 GHz in our final epoch and the spectral index of 0.7 indicate that the radio emission could be due to star formation in the host galaxy. We therefore estimated AT 2020mot's host-galaxy star formation rate (SFR) using infrared photometry in the AllWISE Source Catalog, based on observations by the Wide-field Infrared Survey Explorer [WISE, \cite{2010AJ....140.1868W}]. 
We adopt the host galaxy 95\% confidence upper limit WISE band-4 (22~$\rm\mu$m) magnitude $m_{W4} = 8.944$ mag. We calculate an upper limit host galaxy’s $\rm SFR = 0.08 ~ M_\odot ~ yr^{-1}$ by converting the WISE band-4 magnitude to SFR for z=0.07 [\cite{2017ApJ...850...68C}, their equation 6]. Using the most recent calibration of the radio luminosity - SFR relation \cite{kouroumpatzakis2021} and the observed radio 1.4~GHz flux density $\rm f_{\rm 1.4 GHz} = 85 \pm 25 \mu {\rm Jy}$, we calculate SFR = $\rm 0.27 \pm 0.08 ~ M_\odot~yr^{-1}$. Our calculation did not take into account the intrinsic scatter in the radio-SFR relation, determined by the logarithm of the ratio of the radio ($L_{\rm 1.4~GHz}$) to the $H\alpha$ ($L_{\rm H\alpha}$) luminosity, which can be large [$\rm \delta [{\rm log} L_{\rm 1.4~GHz}/L_{\rm H\alpha}] = 0.52$, \cite{kouroumpatzakis2021}], and already these values are in agreement within 2$\sigma$. We discuss more constrains on the SFR of the host galaxy below.

\begin{table}
\caption{{\bf Radio observations for AT~2020mot}. The first column shows the date of the observations. Columns two and three show the observing band and its corresponding frequency respectively. The last two columns show the estimated flux density and the beam size used. The flux density uncertainty estimates in column four include only the statistical uncertainty. }
\label{tab:radio_obs}
\begin{center}
\begin{tabular}{lcccc}
        \hline
         Date & Band & Mean Frequency & Flux density & Beam size \\
         (MJD) &  & (GHz) & ($\mu$Jy) & (arcsec)\\
         \hline 
         59077.86472 & $K$u & 15 & $<27$ & $0.71\times0.32$ \\
         \hline
         59243.40170 & $C$ & 5 & $87\pm12$ & $0.61\times0.26$ \\
         59243.40170 & $C$ & 7 & $36\pm9$ & $0.45\times0.20$ \\
         \hline
         59273.25564 & $L$ & 1.4 & $86\pm30$ & $1.79\times0.83$ \\
         59273.23174 & $S$ & 2.5 & $103\pm25$ & $1.12\times0.52$ \\
         59273.23174 & $S$ & 3.5 & $78\pm16$ & $0.84\times0.39$ \\
         59273.20547 & $C$ & 5 & $43\pm9$ & $0.56\times0.26$ \\
         59273.20547 & $C$ & 7 & $36\pm11$ & $0.41\times0.19$ \\
         \hline\hline
\end{tabular}
\end{center}
\end{table}

\subsubsection*{Optical and UV Observations}

We analysed twelve \textit{Swift}/Ultra-Violet Optical Telescope (UVOT) observations, taken during the AT 2020mot decay with six optical/UV bands ($U/V/B/W1/M2/W2$), and one short $U$-band observation that included the host galaxy in the field-of-view taken before the TDE. The source count rates, AB magnitudes and fluxes (F) were measured with {\tt uvotsource} task that is part of the High Energy Astrophysics Software (\textsc{HEASoft}) version 6.26.1. If more than one exposure were taken during the telescope pointing, we summed the images using {\tt uvotimsum} task. We also checked that the measured count rates were not affected by the source falling in a low throughput area of the detector.  

We also analysed three \textit{XMM-Newton}/Optical Monitor (OM) observations with $U-$ and $B-$bands taken over 200 days after the peak of the TDE. The source count rates were measured with {\tt omichain} task that is part of the XMM Scientific Analysis System (\textsc{SAS}, \url{https://www.cosmos.esa.int/web/xmm-newton/sas}) version 19.0.0. We converted the source count rates to AB magnitudes and fluxes using the zero points of the OM AB magnitude system and corresponding flux conversion factors.  

\begin{figure}
\centering
\includegraphics[scale=0.7]{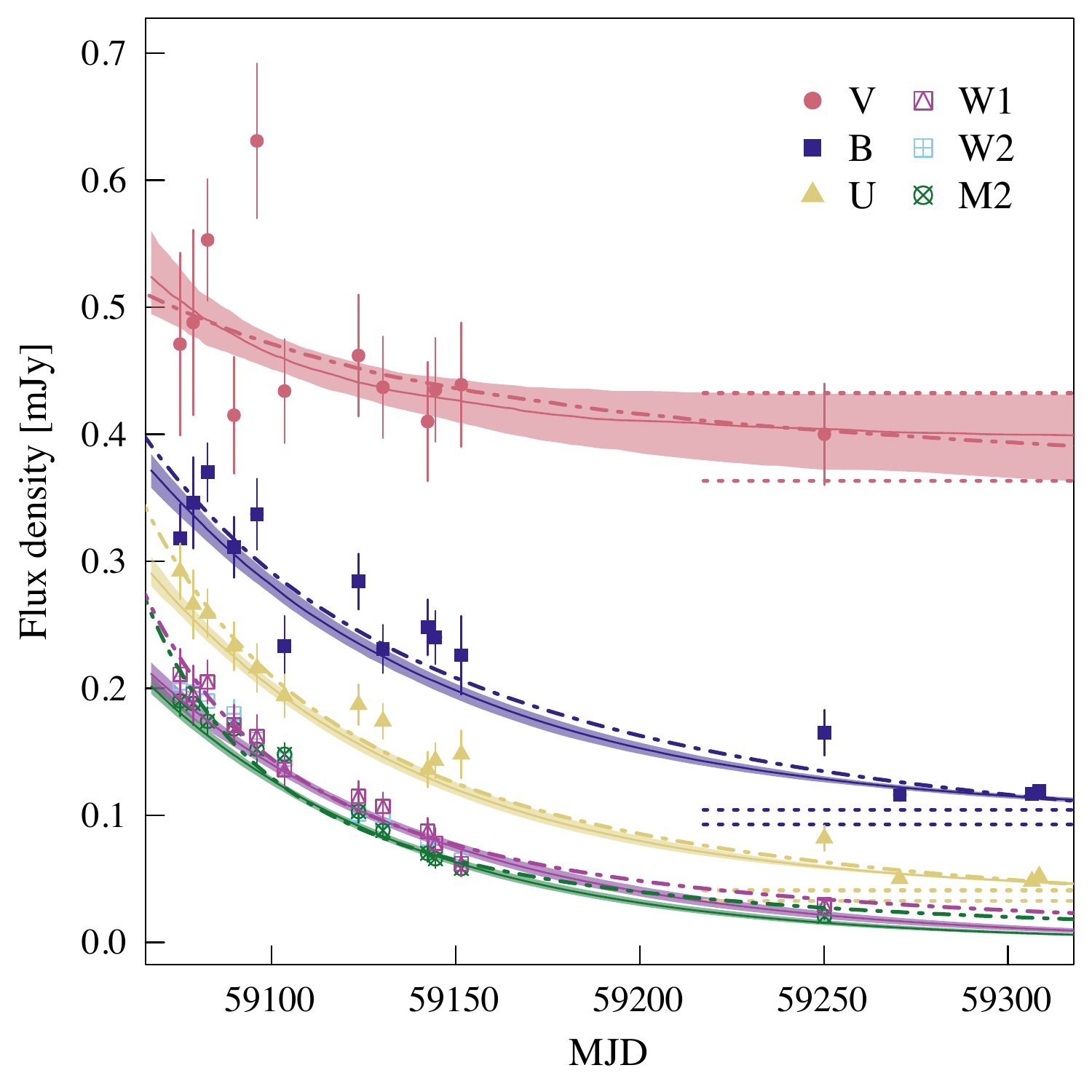}
\caption{{\bf Optical and UV light curves of AT~2020mot from  \textit{Swift}/UVOT and \textit{XMM}/OM.} Different symbols correspond to different optical/UV bands as indicated in the figure's legend. We fit both an exponential decay model (solid lines) and $t^{-5/3}$ model (dot-dashed lines). The shaded areas indicate the 1$\sigma$ confidence levels of the exponential decay models. For the optical bands ($UBV$), the dotted horizontal lines indicate the constant host galaxy contribution (1$\sigma$-confidence range). The galaxy flux at the UV band is negligible. The errorbars indicate the 68\% confidence interval.} 
\label{fig:uv_light}
\end{figure}

All UV observations with the corresponding magnitudes and fluxes are tabulated in Table \ref{tab:uv_obs} and plotted in Fig. \ref{fig:uv_light}. We fitted the decaying fluxes of the AT 2020mot emission with an exponential or $t^{-5/3}$ decay model, using the \textsc{UltraNest} software \cite{buchner21}. \textsc{UltraNest} calculates the posterior probability distributions and the Bayesian evidence using the nested sampling Monte Carlo algorithm MLFriends \cite{buchner14,buchner19}. Figure \ref{fig:uv_light} shows the best-fitting exponential functions ($F \propto e^{(-t/b)}$) with the decay timescales $b$ = 56$\pm$27 days ($V$-band), $b$ = 84$\pm$9 days ($B$-band), $b$ = 75$\pm$8 days ($U$-band), $b$ = 81$\pm$6 days ($W1$-band), $b$ = 71$\pm$4 days ($M2$-band), and $b$ = 74$\pm$3 days ($W2$-band). The combined mean of all decay rates is 75$\pm$2 days. For $V$-, $B$-, and $U$-band fits we add also a constant to the decay model to represent the host galaxy contribution. These resulted in host galaxy fluxes 0.40$\pm$0.03 mJy, 0.099$\pm$0.006 mJy, and 0.037$\pm$0.004 mJy for $V$-, $B$-, and $U$-band, respectively. These correspond to magnitudes 17.4$\pm$0.1, 18.9$\pm$0.1, and 20.0$\pm$0.1. We assume the host galaxy contribution in the UV bands is negligible. The pre-TDE $U$-band flux of 0.05$\pm$0.01 mJy is within 1$\sigma$ of the fitted value. 

In addition to the exponential model, we fitted the TDE decay with the expected $F \propto (t-t_{D})^{-5/3}$ model. The best-fitting functions are shown in Fig. \ref{fig:uv_light} with the disruption time of the star as $t_{D}$ = MJD 58881$\pm$50 ($V$-band), $t_{D}$ = MJD 58924$\pm$22 ($B$-band), $t_{D}$ = MJD 58968$\pm$4 ($U$-band), $t_{D}$ = MJD 58992$\pm$8 ($W1$-band), $t_{D}$ = MJD 59004$\pm$4 ($M2$-band), and $t_{D}$ = MJD 59000$\pm$4 ($W2$-band). The combined mean of the disruption time in the UV bands is MJD 59001$\pm$3 days. The peak emission is achieved at approximately MJD~59077, so this implies $t_{0}\sim$50 days, if $t_{\mathrm{peak}}-t_{D} \sim 1.5 \times t_{0}$ \cite{Ryu2020}. However, the best-fitting values for the optical bands have the disruption time at progressively earlier times. The fitted host galaxy fluxes are very low with 0.006$\pm$0.004 mJy, 0.05$\pm$0.01 mJy, and 0.35$\pm$0.04 mJy for $U-$, $B-$, and $V-$band, respectively. The fitted host galaxy $U-$band flux is more than 4$\sigma$ below the pre-TDE $U-$band flux of 0.05$\pm$0.01 mJy. On the other hand, for the exponential decay model (Fig. \ref{fig:uv_light}) the fitted host galaxy fluxes are close to the magnitudes observed with NOT (see below) that are consistent with the pre-TDE values measured by Pan-STARRS. Therefore, we consider the exponential decay model more likely than the $F \propto (t-t_{D})^{-5/3}$ model, at least in the optical bands.

\begin{table}
\caption{{\bf {\it Swift} and {\it XMM} UV observations of AT~2020mot}. First three columns show the date, and the start, and end time of the observations in Coordinated Universal Time (UTC). Columns four and five show the UV band, and the last two columns the magnitude in the AB system and the corresponding flux density.}
\label{tab:uv_obs}
\begin{center}
\begin{tabular}{lllllll}
         \hline
         Date & Obs. start & Obs. end & Band & Exp. (s) & AB mag & Flux density (mJy)\\
         \hline 
         2019-04-28 & 08:24:24 & 08:25:43 & $U$ & 78 & 19.64$\pm$0.24 & 0.05$\pm$0.01 \\
         \hline
         2020-08-14 & 03:18:07 & 03:21:26 &$W1$& 197 & 18.08$\pm$0.09 & 0.21$\pm$0.02 \\
          & 03:21:31 & 03:23:11 & $U$ & 98 & 17.75$\pm$0.09 & 0.29$\pm$0.02 \\
          & 03:23:15 & 03:24:55 & $B$ & 98 & 17.63$\pm$0.11 & 0.32$\pm$0.03 \\
          & 03:25:01 & 03:31:40 & $W2$ & 392 & 18.13$\pm$0.08 & 0.20$\pm$0.01 \\
          & 03:31:45 & 03:33:25 &$V$& 98 & 17.21$\pm$0.17 & 0.47$\pm$0.07 \\
          & 03:33:29 & 03:38:55 &$M2$& 321 & 18.21$\pm$0.09 & 0.19$\pm$0.01 \\
         \hline
         2020-08-17 & 17:07:35 & 17:09:15 &$W1$& 98 & 18.16$\pm$0.12 & 0.20$\pm$0.02 \\
         & 17:09:20 & 17:10:09 & $U$ & 49 & 17.85$\pm$0.11 & 0.27$\pm$0.03 \\
         & 17:10:15 & 17:11:05 & $B$ & 49 & 17.54$\pm$0.13 & 0.35$\pm$0.04 \\
         & 17:11:10 & 17:14:31 & $W2$ & 198 & 18.16$\pm$0.09 & 0.20$\pm$0.02\\
         & 17:14:36 & 17:15:25 &$V$& 49 & 17.17$\pm$0.17 & 0.49$\pm$0.07 \\
         & 17:15:30 & 17:17:55 &$M2$& 143 & 18.22$\pm$0.11 & 0.19$\pm$0.02 \\
         \hline
         2020-08-21 & 15:04:26 & 15:08:59 &$W1$& 268 & 18.10$\pm$0.09 & 0.21$\pm$0.02 \\
          & 15:09:03 & 15:11:20 & $U$ & 135 & 17.88$\pm$0.08 & 0.26$\pm$0.02\\
          & 15:11:25 & 15:13:41 & $B$ & 135 & 17.47$\pm$0.09 & 0.37$\pm$0.02 \\
          & 15:13:47 & 15:22:54 & $W2$ & 538 & 18.20$\pm$0.07 & 0.19$\pm$0.01 \\
          & 15:23:00 & 15:25:17 &$V$& 135 & 17.04$\pm$0.10 & 0.55$\pm$0.05 \\
          & 15:25:21 & 15:30:56 &$M2$& 329 & 18.31$\pm$0.10 & 0.17$\pm$0.01 \\
         \hline
         2020-08-28 & 19:00:56 & 19:04:39 &$W1$& 220 & 18.30$\pm$0.10 & 0.17$\pm$0.02\\
          & 19:04:44 & 19:06:36 & $U$ & 110 & 18.00$\pm$0.09 & 0.23$\pm$0.02 \\
          & 19:06:40 & 19:08:32 & $B$ & 110 & 17.66$\pm$0.10 & 0.31$\pm$0.02 \\
          & 19:08:38 & 19:16:05 & $W2$ & 440 & 18.26$\pm$0.07 & 0.18$\pm$0.01 \\
          & 19:16:10 & 19:18:02 &$V$& 110 & 17.35$\pm$0.13 & 0.42$\pm$0.05 \\
          & 19:18:06 & 19:23:55 &$M2$& 344 & 18.34$\pm$0.09 & 0.17$\pm$0.01 \\

 \hline
         2020-09-04 & 00:47:00 & 00:49:59 &$W1$& 176 & 18.36$\pm$0.11 & 0.16$\pm$0.02 \\
          & 00:50:03 & 00:51:33 & $U$ & 88 & 18.08$\pm$0.10 & 0.22$\pm$0.02 \\
          & 00:51:38 & 00:53:08 & $B$ & 88 & 17.57$\pm$0.11 & 0.34$\pm$0.03 \\
          & 00:53:13 & 00:59:11 & $W2$ & 352 & 18.39$\pm$0.08 & 0.16$\pm$0.01 \\
          & 00:59:16 & 01:00:46 &$V$& 88 & 16.89$\pm$0.11 & 0.63$\pm$0.06 \\
          & 01:00:50 & 01:05:55 &$M2$& 300 & 18.45$\pm$0.10 & 0.15$\pm$0.01 \\

         \hline\hline
         (continued on next page)
\end{tabular}
\end{center}
\end{table}

\setcounter{table}{2}

\begin{table}
\caption{\bf Continued.}
\begin{center}
\begin{tabular}{lllllll}
         \hline
         Date & Obs. start & Obs. end & Band & Exp. (s) & AB mag & Flux (mJy)\\         
    
         \hline
        
         2020-09-11 & 12:31:52 & 12:35:02 &$W1$& 315 & 18.55$\pm$0.10 & 0.14$\pm$0.01\\
          & 12:35:07 & 12:36:41 & $U$ & 157 & 18.20$\pm$0.10 & 0.19$\pm$0.02 \\
          & 12:36:46 & 12:38:21 & $B$ & 157 & 17.97$\pm$0.13 & 0.23$\pm$0.02 \\
          & 12:38:26 & 12:44:45 & $W2$ & 627 & 18.50$\pm$0.07 & 0.14$\pm$0.01 \\
          & 12:44:50 & 12:46:25 &$V$& 157 & 17.30$\pm$0.11 & 0.43$\pm$0.04 \\
          & 12:46:30 & 12:51:55 &$M2$& 553 & 18.48$\pm$0.09 & 0.15$\pm$0.01 \\          
         
          \hline
         2020-10-01 & 14:28:36 & 14:32:54 &$W1$& 254 & 18.73$\pm$0.11 & 0.12$\pm$0.01 \\
          & 14:32:58 & 14:35:07 & $U$ & 127 & 18.24$\pm$0.09 & 0.19$\pm$0.02 \\
          & 14:35:12 & 14:37:21 & $B$ & 127 & 17.76$\pm$0.10 & 0.28$\pm$0.02 \\
          & 14:37:26 & 14:46:03 & $W2$ & 509 & 18.89$\pm$0.08 & 0.101$\pm$0.007 \\
          & 14:46:08 & 14:48:17 &$V$& 127 & 17.23$\pm$0.12 & 0.46$\pm$0.05 \\
          & 14:48:22 & 14:54:55 &$M2$& 388 & 18.87$\pm$0.11 & 0.10$\pm$0.01 \\
          \hline
         2020-10-08 & 05:56:27 & 05:58:15 &$W1$& 322 & 18.82$\pm$0.10 & 0.11$\pm$0.01 \\
          & 05:58:20 & 05:59:13 & $U$ & 161 & 18.32$\pm$0.09 & 0.17$\pm$0.01 \\
          & 05:59:18 & 06:00:12 & $B$ & 161 & 17.98$\pm$0.10 & 0.23$\pm$0.02 \\
          & 06:00:17 & 06:03:52 & $W2$ & 643 & 18.94$\pm$0.08 & 0.097$\pm$0.007 \\
          & 06:03:57 & 06:04:50 &$V$& 161 & 17.29$\pm$0.11 & 0.44$\pm$0.04 \\
          & 06:04:54 & 06:06:55 &$M2$& 424 & 19.04$\pm$0.11 & 0.088$\pm$0.007 \\
         \hline
         2020-10-20 & 08:55:46 & 08:59:30 &$W1$& 220 & 19.03$\pm$0.13 & 0.09$\pm$0.01 \\
          & 08:59:35 & 09:01:26 & $U$ & 110 & 18.59$\pm$0.12 & 0.14$\pm$0.01 \\
          & 09:01:31 & 09:03:23 & $B$ & 110 & 17.90$\pm$0.11 & 0.25$\pm$0.02 \\
          & 09:03:29 & 09:10:56 & $W2$ & 441 & 19.19$\pm$0.10 & 0.076$\pm$0.007 \\
          & 09:11:01 & 09:12:53 &$V$& 110 & 17.36$\pm$0.13 & 0.41$\pm$0.05 \\
          & 09:12:58 & 09:18:57 &$M2$& 354 & 19.29$\pm$0.13 & 0.070$\pm$0.007 \\
         \hline
         2020-10-22 & 11:51:05 & 11:53:41 &$W1$& 153 & 19.16$\pm$0.16 & 0.08$\pm$0.01 \\
          & 11:53:46 & 11:55:03 & $U$ & 157 & 18.53$\pm$0.11 & 0.14$\pm$0.01 \\
          & 11:55:08 & 11:56:26 & $B$ & 157 & 17.94$\pm$0.11 & 0.24$\pm$0.02 \\
          & 11:56:32 & 12:01:43 & $W2$ & 307 & 19.25$\pm$0.12 & 0.073$\pm$0.007 \\
          & 12:01:48 & 12:03:06 &$V$& 157 & 17.30$\pm$0.11 & 0.44$\pm$0.04 \\
          & 12:03:11 & 12:07:55 &$M2$& 280 & 19.36$\pm$0.15 & 0.066$\pm$0.008 \\
         \hline
         2020-10-29 & 12:36:05 & 12:39:43 &$W1$& 215 & 19.41$\pm$0.17 & 0.06$\pm$0.01 \\
          & 12:39:48 & 12:41:37 & $U$ & 107 & 18.49$\pm$0.15 & 0.15$\pm$0.02 \\
          & 12:41:41 & 12:43:30 & $B$ & 107 & 18.00$\pm$0.17 & 0.23$\pm$0.03 \\
          & 12:43:36 & 12:50:53 & $W2$ & 431 & 19.37$\pm$0.11 & 0.065$\pm$0.006 \\
          & 12:50:58 & 12:52:47 &$V$& 107 & 17.29$\pm$0.13 & 0.44$\pm$0.05 \\
          & 12:52:51 & 12:58:55 &$M2$& 358 & 19.49$\pm$0.14 & 0.058$\pm$0.006 \\
         \hline \hline
         (continued on next page)
\end{tabular}
\end{center}
\end{table}

\setcounter{table}{2}

\begin{table}
\caption{\bf Continued.}
\begin{center}
\begin{tabular}{lllllll}
         \hline
         Date & Obs. start & Obs. end & Band & Exp. (s) & AB mag & Flux (mJy)\\         
          \hline         
         2021-02-05 & 02:36:43 & 02:40:43 &$W1$& 600 & 20.21$\pm$0.16 & 0.030$\pm$0.005 \\
          & 02:40:47 & 02:42:47 & $U$ & 299 & 19.14$\pm$0.14 & 0.08$\pm$0.01 \\
          & 02:42:52 & 02:44:52 & $B$ & 299 & 18.34$\pm$0.13 & 0.17$\pm$0.02 \\
          & 02:44:58 & 02:52:58 & $W2$ & 1200 & 20.42$\pm$0.12 & 0.025$\pm$0.003 \\
          & 02:53:03 & 02:55:03 &$V$& 299 & 17.39$\pm$0.12 & 0.40$\pm$0.04 \\
          & 02:55:07 & 03:01:55 &$M2$& 1002 & 20.63$\pm$0.16 & 0.021$\pm$0.003 \\
         \hline
         2021-02-25 & 12:19:21 & 13:37:47 & $U$ & 4400 & 19.66$\pm$0.04 & 0.050$\pm$0.002 \\
          & 13:37:48 & 14:56:14 & $B$ & 4400 & 18.74$\pm$0.03 & 0.116$\pm$0.003 \\
         \hline
         2021-04-02 & 10:16:37 & 11:35:03 & $U$ & 4400 & 19.69$\pm$0.04 & 0.048$\pm$0.002 \\
          & 11:35:04 & 12:53:30 & $B$ & 4400 & 18.73$\pm$0.03 & 0.117$\pm$0.003 \\
         \hline
         2021-04-04 & 10:08:33 & 11:25:59 & $U$ & 4400 & 19.61$\pm$0.04 & 0.052$\pm$0.002 \\
          & 11:27:00 & 12:45:26 & $B$ & 4400 & 18.71$\pm$0.03 & 0.119$\pm$0.003 \\
         \hline\hline
\end{tabular}
\end{center}
\end{table}
\newpage

\begin{figure}
\centering
\includegraphics[scale=0.6]{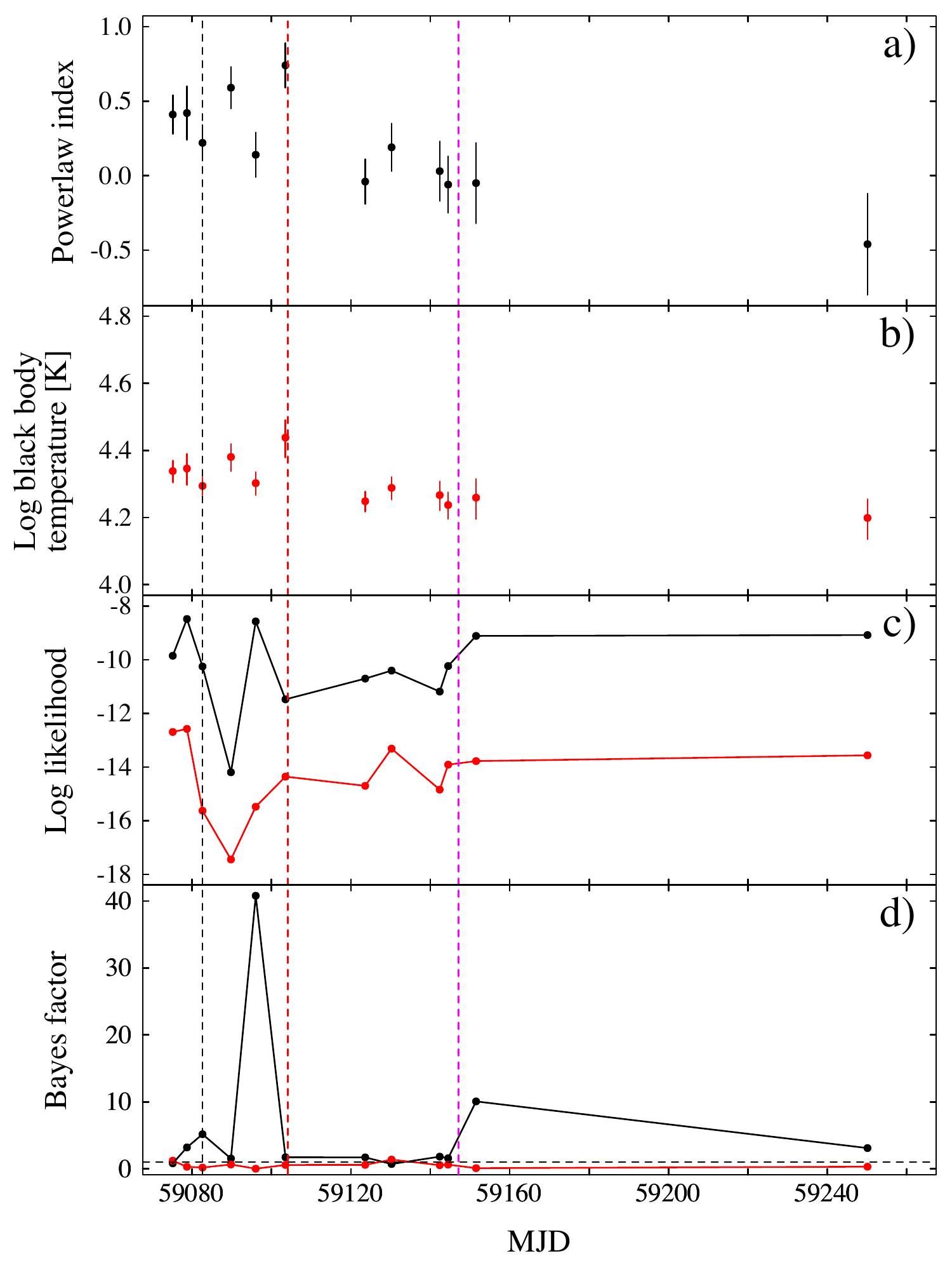}
\caption{{\bf Model parameters from the power-law (black points and lines) and black body (red points and lines) models fitted to the UVOT SEDs}. The figure panels show the evolution of the power-law index (a), the black body temperature (b), the log likelihood (c), and the Bayes factor (d). The black, red, and magenta dashed lines show the location of 1.5$t_{0}$, 2$t_{0}$ and 3$t_{0}$ as in Fig. \ref{fig:lightcurve}. The errorbars indicate the 68\% confidence interval.} 
\label{fig:mod_param}
\end{figure}

\begin{figure}
\centering
\includegraphics[scale=0.8]{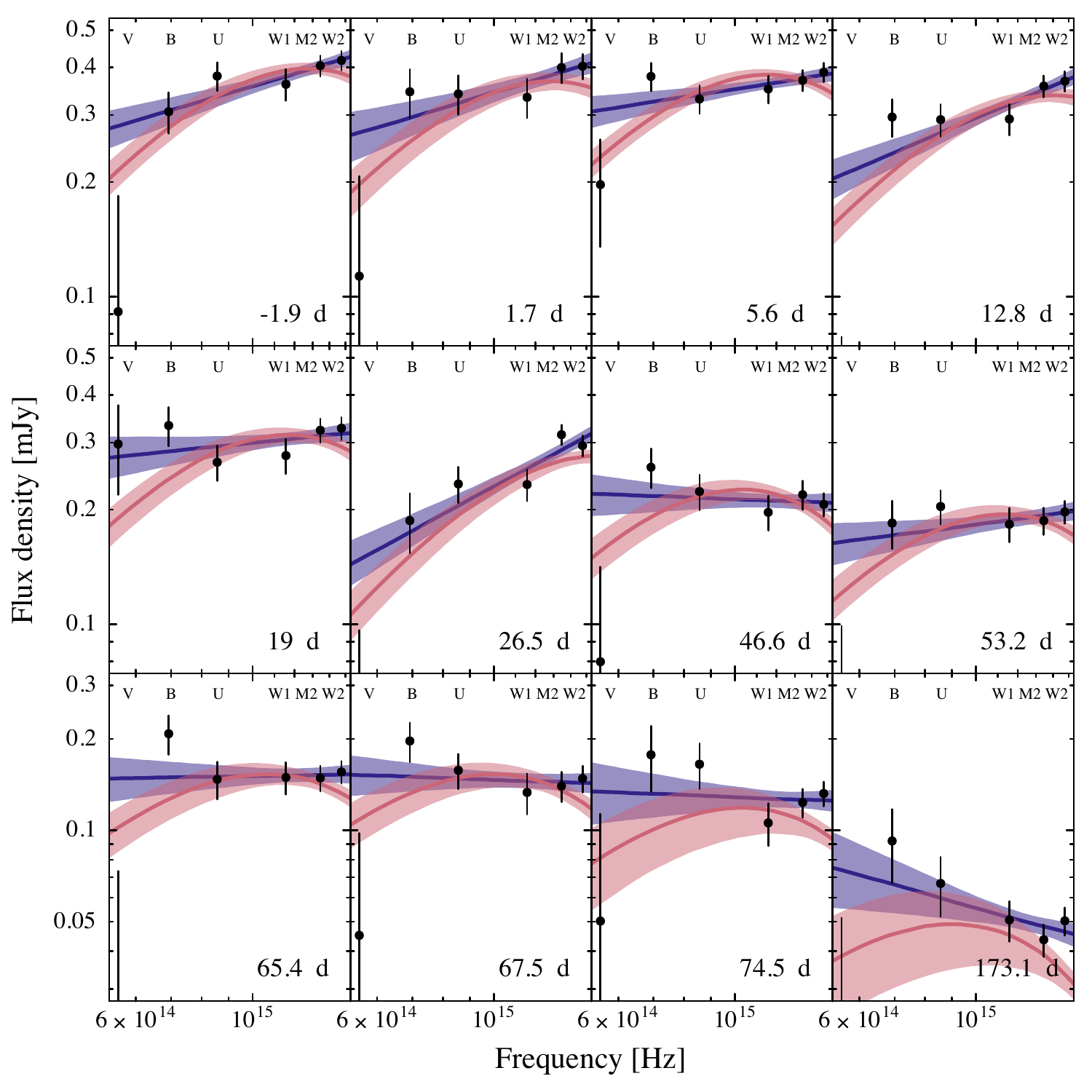}
\caption{{\bf Dereddened and host galaxy subtracted \textit{Swift}/UVOT SEDs from all observing epochs}. All epochs were fitted with a power-law model (blue) and a black body model (red). The blue and red shaded regions show the 1$\sigma$ uncertainty for the power-law and black body models respectively. The time in each panel is measured from the estimated peak date at MJD 59077. For the power-law model, the spectra are optically thick between $t_{0}$ and 2$t_{0}$, approximately flat between 2$t_{0}$ and 3$t_{0}$, and optically thin at late time (see also Fig. \ref{fig:mod_param}). The optical polarization observations were taken approximately 24, 46, 77, and 87 days after the peak. The errorbars indicate the 68\% confidence interval.} 
\label{fig:dered_sed}
\end{figure}

To study the UV spectral evolution during the AT 2020mot decay, we subtracted the host galaxy contribution from the $V-$, $B-$, and $U-$bands determined from fitting the decay curves, and dereddened the fluxes using the standard extinction [R($V$)] profile \cite{fitzpatrick99}, R($V$) = 3.1, and $E(B-V) = 0.088$ that we estimated from the SDSS $g$- and $r$-band extinction [$g$ - $r$ = 0.287 mag - 0.199 mag, \cite{schlafly11}]. The resulting host-galaxy-subtracted and dereddened spectral energy distributions (SEDs) were fitted with both a power-law and a black body model. Figure \ref{fig:mod_param} shows the model parameter evolution, log-likelihood, and the Bayes factors of the two models. In the majority of cases, we cannot differentiate between the two models based on the Bayes factor. In five cases, the power-law model is preferred, with Bayes factor showing substantial evidence ($>$3), and very strong evidence [$\sim$40] in one case. Typically in TDEs, the black body temperature is approximately constant \cite{Hung17} or has a small increase towards the end of the event \cite{vanVelzen21}, neither which is the case here. In Fig. \ref{fig:dered_sed}, we show the SEDs together with the best-fitting models. The best-fitting power-law models are optically thick (power-law indices ranging from 0.2 to 0.8) at the beginning of the decay from 1.5$t_{0}$ to 2$t_{0}$ (i.e. when the nozzle shock is still dominating), flat (power-law index $\sim$0) during 2$t_{0}$ to 3$t_{0}$ (i.e. when the reverse and forward shocks start to dominate), and optically thin (power-law index $<$0) at late times. We interpret this as due to the nozzle shock region closer to the black hole having a larger optical depth than the forward and reverse shock regions.

\subsubsection*{X-ray Observations}

\begin{figure}
\centering
\includegraphics[scale=0.4]{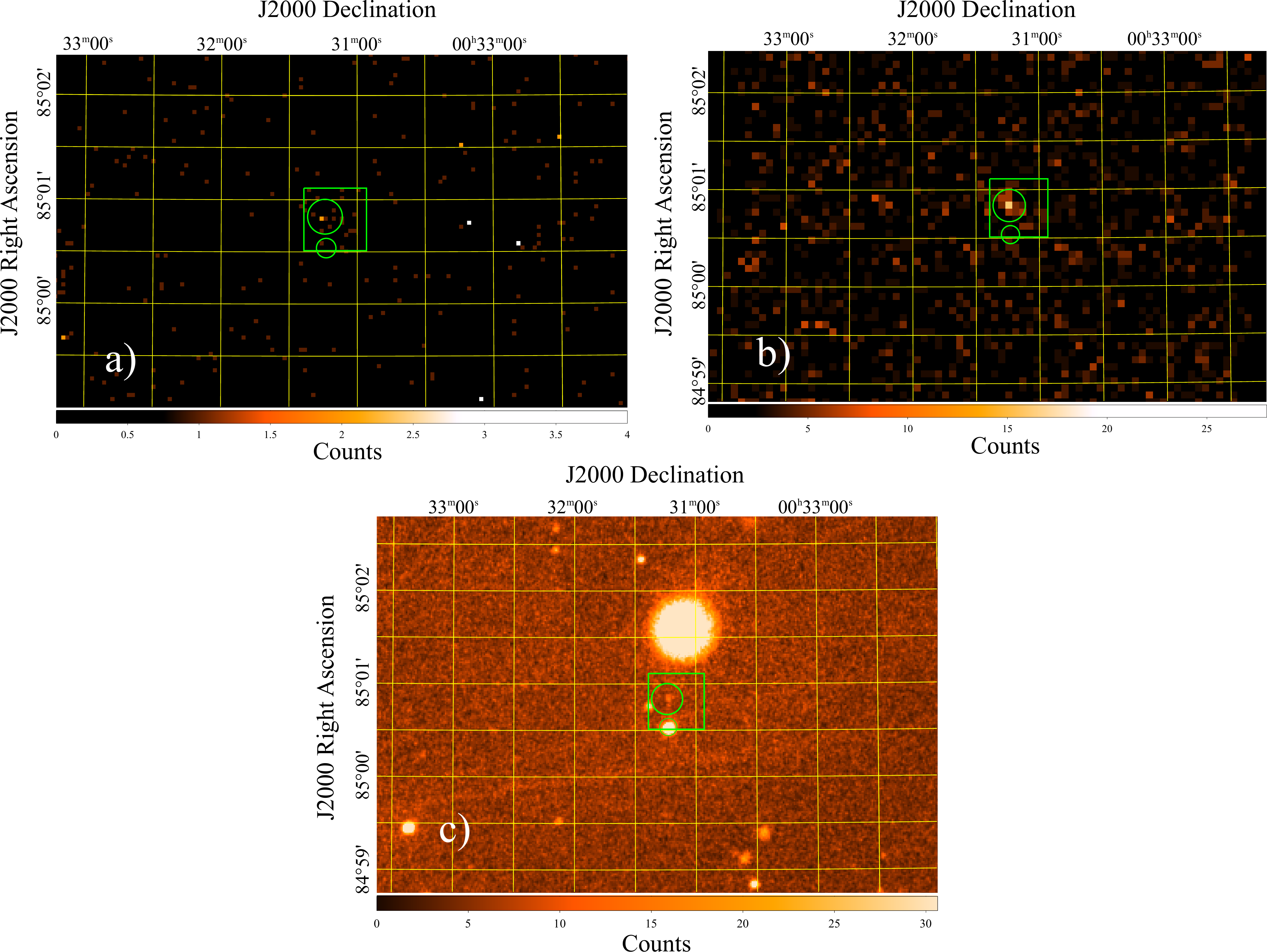}
\caption{{\bf X-ray field of view from the direction of AT~2020mot from \textit{Swift} and \textit{XMM}}. \textit{(a)} stacked \textit{Swift}/XRT image. \textit{(b)} stacked \textit{XMM}/MOS image.  \textit{(c)} stacked \textit{Swift}/UVOT $W2$-band image. All panels show the source region of the XRT source (36$^{\prime\prime}\times36^{\prime\prime}$ box), the source region of the MOS source (10$^{\prime\prime}$ circle), and the standard UVOT source extraction region size (5$^{\prime\prime}$ circle) located at the coordinates of AT 2022mot. The number of counts received is shown in the color bar under the panels.}	
\label{fig:xray_fov}
\end{figure}

We obtained \textit{Swift}'s X-Ray Telescope (XRT) observations simultaneous to the UVOT observations described above, to search for an X-ray counterpart to AT 2020mot. We extracted images from the photon counting mode data of XRT, but no counts were detected at the location of AT 2020mot. Therefore, we searched for sources in a stacked image using \textit{Swift} \textsc{DeepSky} pipeline \cite{Giommi2019}. A 4$\sigma$ source was found at the vicinity of AT 2020mot with a count rate of 0.0016$\pm$0.0004 $\rm counts\cdot s^{-1}$ (Fig. \ref{fig:xray_fov}a).      

Because the XRT source is marginally consistent with the location of AT 2020mot, we observed the source with \textit{XMM-Newton} on three occasions (last three dates in Table \ref{tab:uv_obs}). All pointings were heavily affected by soft proton radiation from the Earth's magnetosphere, thus rendering the European Photon Imaging Camera PN (EPIC-pn) camera data unusable. We therefore used part of the Metal Oxide Semi-conductor (MOS) camera data. We extracted the source images using \textsc{xmmextractor} task in SAS. We modified the configuration file to restrict the MOS background flaring rate (i.e flux in the 10-12 keV band) below 0.35 $\rm counts\cdot s^{-1}$.

We stacked all MOS images using \textsc{multixmmselect} task in SAS resulting in $\sim$16.6 ksec of exposure time. Visually inspecting the stacked image, we did not see a source at the location of AT 2020mot (small circle in Fig. \ref{fig:xray_fov}b). However, about 15$^{\prime\prime}$ north from that location was a visible source that is more central to the XRT source box (Fig. \ref{fig:xray_fov}a). From the stacked image we calculated the count rates at the location of AT 2020mot, the nearby source, and a background region using \textsc{eregionanalyse} task in SAS. The region sizes were 10$^{\prime\prime}$ for the sources and 60$^{\prime\prime}$ for the background. The source count rate was then estimated as $(S-B\times As/Ab)/EEF$, where $S$ is the source region count rate, $B$ is the background region count rate, $As$ is the source region size, $Ab$ is the background region size, and $EEF$ is the encircled energy factor  $EEF=0.8$, given by the {\tt eregionanalyse} routine. The corresponding uncertainty is calculated as $\sqrt{S + B\times(As/Ab)^2}$. The resulting 2$\sigma$ upper limit for the count rate in the location of AT 2020mot is $<$0.002 $\rm counts\cdot s^{-1}$, while for the nearby source it is 0.0025$\pm$0.0006 $\rm counts\cdot s^{-1}$. Using {\tt WebPIMMS} (based on HEASARC tool PIMMS v.4.11b), we converted the XRT Photon-Counting (PC) count rate to the expected value for the thin \textit{XMM}/MOS filter, $\sim$0.004 $\rm counts\cdot s^{-1}$, which corresponds to about the same count rate as observed in the MOS nearby source region. We stacked all the UVOT images from $W2$-band (the highest UVOT frequency) and found a faint source at the location of the nearby \textit{XMM} source. Figure \ref{fig:xray_fov} shows the location of the nearby source as a 10$^{\prime\prime}$ circle, while the location of AT 2022mot is shown as a smaller 5$^{\prime\prime}$ circle. These are clearly two different sources as can be seen in the XMM/MOS (Fig. \ref{fig:xray_fov}b) and Swift/UVOT (Fig. \ref{fig:xray_fov}c) images. Therefore, we attribute all the X-ray emission to the nearby source.

Assuming thermal accretion disk emission with a temperature of 60 eV \cite{Shu20} and a line-of-sight Galactic hydrogen column density of $N_{H}=5.8 \times 10^{20}$ atoms cm$^{-2}$ \cite{HI4PI}, the \textit{XMM}/MOS 2$\sigma$ upper limit of 0.002 $\rm counts\cdot s^{-1}$ corresponds to an accretion disk luminosity of $L_{X} < 1.2 \times 10^{41}$ $\rm erg\cdot s^{-1}$ at the redshift of AT 2020mot in the 0.3--10 keV band that corresponds to an X-ray luminosity over the Eddington luminosity ($L_{Edd}$) of  $L_{X}/L_{Edd} \sim 2.6 \times 10^{-4}$ for a black hole mass $M_{BH}=3.6\times10^{6} \ M_{\odot}$. To compare the optical/X-ray emission ratios to other TDEs (see below), we take the black body model from the optical peak and estimate its bolometric luminosity as $L_{bb}=(1\pm0.2)\times 10^{44}$ $\rm erg\cdot s^{-1}$. Therefore, the optical/X-ray emission ratio of AT 2020mot is higher than $L_{bb}/L_{X} \gtrsim 1000$. All the previously observed optically bright TDEs with detected X-ray emission have $L_{bb}/L_{X} < 1000$ \cite{vanVelzen21,Wevers2019}. This suggests that the accretion disk emission in AT 2020mot has not formed or is completely obscured.

\subsubsection*{Host galaxy and black hole mass characterization}

\begin{figure}
\centering
\includegraphics[scale=0.6]{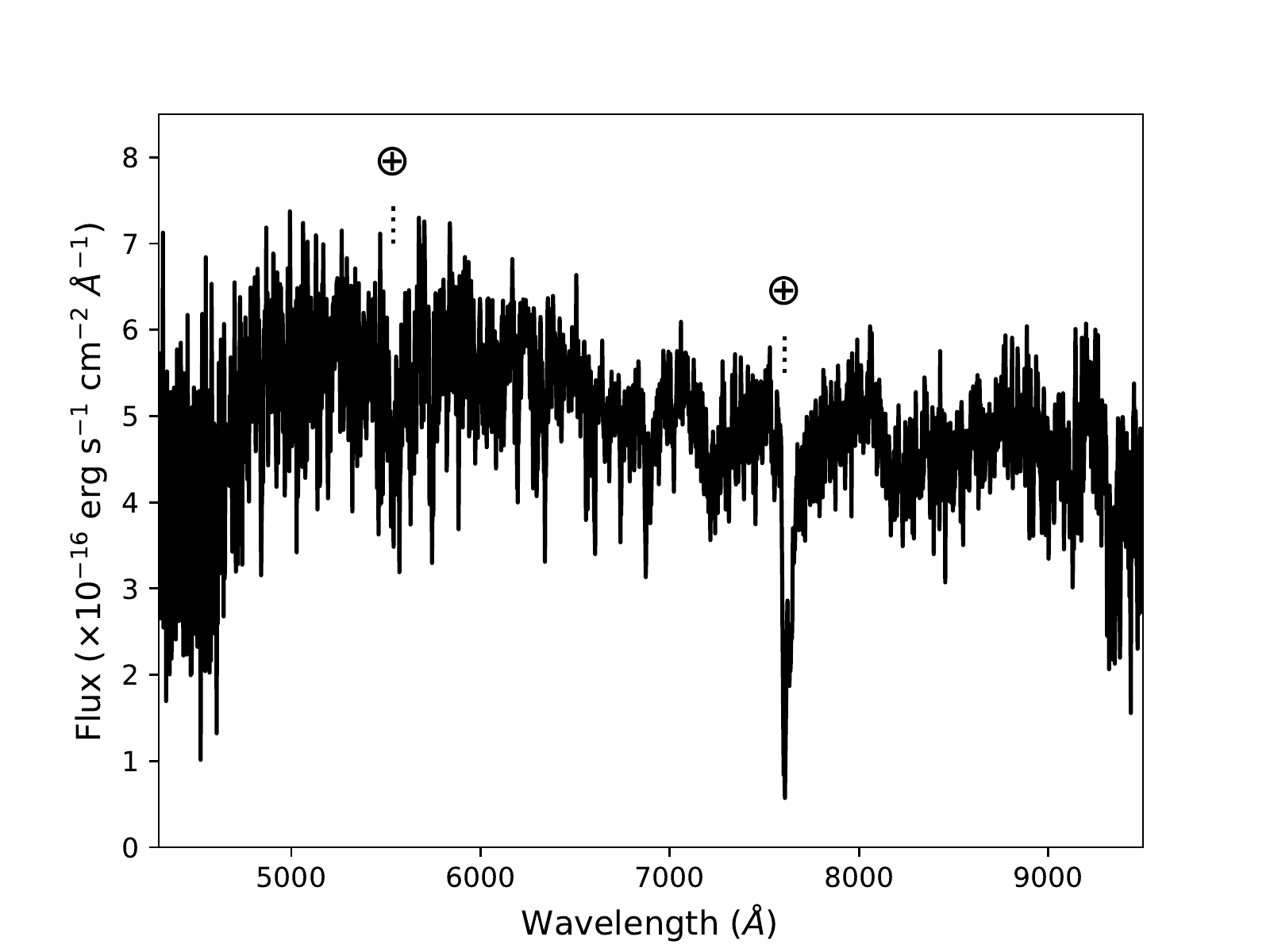}
\caption{ {\bf Optical spectrum of AT 2020mot.} The specturm was taken using ALFOSC on the 19 February 2021 (MJD~59264). Telluric absorption lines are marked with the $\oplus$ symbol and dotted lines.}
\label{fig:not_spec}
\end{figure}

To characterize the host galaxy, estimate its contribution to the polarization and flux measurements, and estimate the black hole mass we performed  imaging and spectroscopic observations of AT 2020mot on 19 February 2021 (MJD~59264) using ALFOSC at the Nordic Optical Telescope. The seeing conditions during our observations were between 1$^{\prime\prime}$ to 2$^{\prime\prime}$. The imaging observations were performed in $U,B,V,R$ and $i$ bands. The observations were analyzed using standard procedures of differential photometry with a semi-automatic pipeline \cite{Nilsson2018}. The comparison and control star (Cl* NGC 188 DGV 325 and Cl* NGC 188 PKM 7436) magnitudes were obtained from Pan-STARRS using filter transformations of \cite{2012ApJ...750...99T} in $B$, $V$, $R$ and $i$ bands and the {\it Swift}-UVOT data in $U-$band. The spectroscopic data was obtained using grisms \#7 (3650-7110 \AA) and \#20 (5650-10150 \AA). The exposure time was 1800s for grism \#7, and 900s for grism \#20. We carried out standard data reduction with \textsc{IRAF} \cite{IRAF}. The flux of the spectrum was cross-calibrated using the magnitudes derived from our quasi-simultaneous photometric observations. The resulting spectrum is shown in Fig. \ref{fig:not_spec}.

The measured magnitudes in different bands are given in Table \ref{tab:phot}. They are all consistent with Pan-STARRS pre-TDE magnitudes, i.e. with originating solely from the host galaxy. However, to measure the host galaxy contribution to $R$-band polarization data (see above), we used the Pan-STARRS pre-TDE values in $g$- and $r$-bands \cite{PANSTARRS_database}, which we convert to  $R=17.0\pm0.1$ or $\rm F{_{\mathrm R}}$ $0.48\pm0.05$\,mJy. The spectrum in Fig. \ref{fig:not_spec} shows no strong emission lines, consistent with the previous classification of the galaxy as an E+A galaxy \cite{2020TNSCR2478....1H}. We do not detect $\rm H{\alpha}$, but as above, this is consistent with the estimated low star-formation rate.

\begin{table}
\caption{{\bf Optical photometric observations of AT~2020mot from the Nordic Optical Telescope}. Each column shows the brightness in magnitudes for a different optical band. The magnitudes are calculated in the AB system. }
\label{tab:phot}
\begin{center}
\begin{tabular}{ccccc}
         \hline
         $U$ & $B$ &$V$& $R$ & $I$ \\
         
         [AB mag] & [AB mag] & [AB mag] & [AB mag] & [AB mag] \\ 
         \hline
         $19.40\pm0.15$&$18.38\pm0.05$&$17.46\pm0.02$&$16.83\pm0.02$&$16.43\pm0.02$ \\ \hline
\end{tabular}
\end{center}
\end{table}

\begin{figure}
\centering
\includegraphics[scale=5.5]{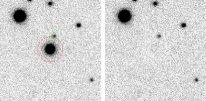}
\caption{{\bf Summed $R$+$i$ band image of the host galaxy}.  The left panel shows the summed $R$ + $i$ image.  The right panel shows the image after subtracting the host galaxy model. The size of the images is 243x243 arcsec. The star in the upper left corner was used as the PSF star. Only pixels within the red circle were included in the fit while the pixels inside the green circle were excluded. }
\label{fig:host_Ri}
\end{figure}

We fitted a Sersic profile \cite{Sersic1968}  to the summed $R$+$i$ band image (Fig. \ref{fig:host_Ri})  following \cite{nilsson03}. The best-fitting effective radius of the galaxy was $1.0\pm0.3^{\prime\prime}$, corresponding to 0.87 kpc at z=0.07. Therefore, we assume that all of the galaxy flux is contained within aperture of 4.5$^{\prime\prime}$ that was used for the analysis of the polarimetric data. The best-fit Sersic index was $4.4\pm0.3$, strongly favoring an elliptical galaxy. We derived an absolute $R-$band magnitude of $M_R=-20.76\pm0.05$. Assuming that all of the luminosity originates from the bulge, this luminosity corresponds to a black hole mass of  $(7\pm3)\times 10^{6}M_\odot$ using a standard relation \cite{2013ApJ...764..151G}. The errorbar includes the uncertainty contribution from the photometry, conversion to the $Ks$-band and the errors
in the $\rm M_{bulge}-M_{BH}$ relation.

\begin{figure}
\centering
\includegraphics[scale=1.5]{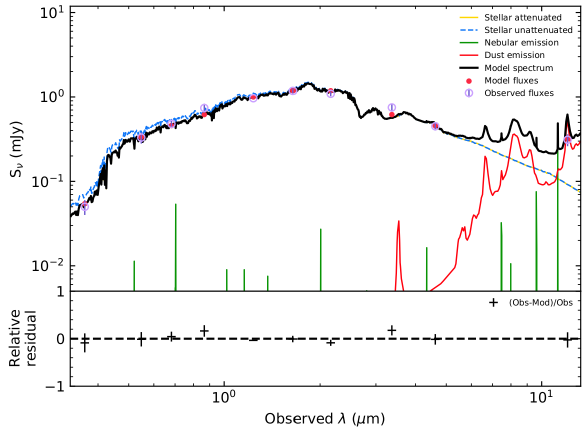}
\caption{{\bf Spectral energy distribution modeling for the host galaxy using \textsc{CIGALE}}. The model is fitted to archival $U$, $V$, $R$, $I$, $J$, $H$, $K_{\rm s}$, WISE bands 1, 2, and 3, pre-TDE magnitudes from Pan-STARRS, 2MASS, and WISE. Observed and model flux densities are shown with the filled-red and open-purple circles respectively. The colored lines show the different model components, while the black solid line shows the best-fit model. The errorbars indicate the 68\% confidence interval. }
\label{fig:host_sed}
\end{figure}

We investigate the SFR of the host galaxy and its consistency with the radio observations (see above). The $\rm H\alpha$ line is the strongest emission line in the visible spectrum, which indicates recent star formation. For our maximum derived star-formation rate $\rm SFR_{1.4~GHz} = 0.27 ~ M_\odot~yr^{-1}$ we estimate the flux of the expected H$\alpha$ line to be $f_{\rm H\alpha} \simeq 5 \times 10^{-15} ~  {\rm erg~s^{-1}}$. Assuming no extinction, that the emission line is not blended with an absorption line, and an undispersed narrow line with full width at half-maximum $\rm FWHM \simeq 10 ~ \AA$, the peak of the emission line would be $F_\nu \simeq 1.5 \times 10^{-16} ~ {\rm erg ~s^{-1}~\AA^{-1}}$. This would be a marginally detectable line, however, if any of the above assumptions  do not hold, the emission line would not be detectable. For example,  for $\rm FWHM \simeq 20 ~ \AA$, the peak of the emission line would be at $f_\nu \simeq 0.9 \times 10^{-17} ~ {\rm erg ~s^{-1}~\AA^{-1}}$ , which is undetectable. We fitted SED of the host galaxy using the \textsc{CIGALE} \cite{2005MNRAS.360.1413B,2009A&A...507.1793N,2019A&A...622A.103B} code based on the available archival photometry from Pan-STARRS [Johnson $U$, $V$, $R$, $I$, \cite{PANSTARRS_database}], the Two Micron All Sky Survey [2MASS, \cite{2MASS_paper}] $J$, $H$, $K_{\rm s}$, and WISE bands 1, 2, and 3 \cite{2010AJ....140.1868W}; WISE-4 was not included because \textsc{CIGALE} does not support upper limits). We applied a standard star-formation history with optional delayed burst and an earlier stellar population, commonly used to characterize a variety of galaxy star-formation histories, while scanning a large parameter space for stellar populations, nebular and dust attenuation values yielding a total of 604,800 models. The model is more than adequate for a passive galaxy like the host of AT~2020mot resulting to a best-fitting model with a reduced $\chi^2=0.39$ (Fig. \ref{fig:host_sed}). More than 10,000 models yielded a reduced $\chi^2<0.5$, hence, we do not consider the low value of the reduced $\chi^2$ as overfitting, rather we attribute it to the passive, thus, easier-to-model, nature of the host galaxy. The galaxy's light is dominated by emission from an old stellar population ($\simeq 5$~Gyr). The SED model fitting indicated $\rm SFR=0.11^{+0.21}_{-0.11}~M_\odot~yr^{-1}$, stellar mass $M_\star = 1.37 \times 10^{10}~{\rm M_\odot}$, metallicity $Z = 0.05$, and low internal reddening $E(B-V)=0.04$~mag, consistent with a gas-and-dust-depleted early-type galaxy with low star-forming activity. Therefore, the lack of strong emission lines, SED, and IR observations, are all consistent with the rather low level of host-galaxy star-forming activity derived from the radio observations.

To constrain the BH mass and mass of the disrupted star we use the \textsc{TDEmass} package \cite{Ryu2020}. \textsc{TDEmass} is based on the outer shock model \cite{Piran2015} and uses the peak luminosity and peak black body temperature. We modeled the closest {\it Swift}-UVOT SED to the outburst peak with a black body model (Fig. \ref{fig:dered_sed}), finding a bolometric peak luminosity of $L_\mathrm{max}= (1\pm0.2)\times10^{44}$ $\rm erg\cdot s^{-1}$ and a temperature of   $T_\mathrm{max}=(2.2\pm{0.2})\times 10^{4}$ K. The peak luminosity was calculated using a Hubble constant of $\rm H_0 = 68~km\cdot s^{-1}\cdot Mpc^{-1}$. This yields $M_\star=1.3^{+0.6}_{-0.3}M_\odot$ and $M_{\rm BH}=3.6^{+1.2}_{-1.4} \times{10^6} M_\odot$, consistent with the upper-limit from the host-galaxy bulge. Using these estimates we calculate [\cite{Ryu2020}, their equations 1, 2 and 5] we calculate the return time of the most bound material to be $t_0= 43^{+16}_{-12}$ days.

\subsubsection*{AT 2020mot compared to other TDEs}

Here we evaluate the rarity of events like AT 2020mot. We use the observed properties of the event which we compare to previous studies of other TDEs.

The host galaxy of AT 2020mot was classified as an E+A type galaxy \cite{2020TNSCR2478....1H}. This host-galaxy type has been found to be the most common type among TDEs \cite{Hammerstein2021}. At a redshift of $z=0.07$ it is near the median redshift of optically selected TDEs [$<z>=0.08$, \cite{VanVelzen2020}]. The median $M_{\rm BH}$ of previous TDEs is about $<M_{\rm BH}>=1.23\times{10^6} M_\odot$ \cite{Wevers2017, Mockler2019, Zhou2021} which is within the uncertainties of our inferred value. We used a Gaussian rise+exponential decay model \cite{vanVelzen21} and the ZTF light curve to estimate the rise time to the peak to be $55\pm5$ days. This places AT 2020mot within 1$sigma$ from the population [$61^{+48}_{-10}$ days, \cite{Mockler2019}] . The lowest peak X-ray luminosity of X-ray detected TDEs is $L_X=5\times10^{41}$ $\rm erg\cdot s^{-1}$ \cite{Saxton2020}, consistent with our upper limit of $L_X<1.2\times10^{41}$ $\rm erg\cdot s^{-1}$ from the {\it XMM-Newton} observation. Our estimated peak temperature of $T_\mathrm{max}=(2.2\pm{0.2})\times 10^{4}$ K is also consistent with the median ($<T_\mathrm{max}>\approx(2.1\pm{0.7})\times 10^{4}$ K) of the population \cite{vanVelzen21}. The estimated peak black body luminosity of $L_\mathrm{bb}=(1\pm{0.2})\times10^{44}$ $\rm erg\cdot s^{-1}$ is within 1$\sigma$ of the population median [$<L_\mathrm{bb}>=1.2\times10^{44}$  $\rm erg\cdot s^{-1}$ , \cite{VanVelzen2020}]. We therefore conclude that AT 2020mot is a typical optically selected TDE. 

There are four other TDEs with detected optical polarization \cite{Wiersema2012,Higgins2019,Lee2020,Wiersema2020}.  {\it Swift} J1644+57 and {\it Swift} J2058.4+0516 (also known as {\it Swift} J2058+0516) are jetted TDEs, OGLE16aaa is an optical TDE, and AT 2019dsg has been interpreted as having a jet  \cite{Stein2021}, or as a non-relativistic outflow \cite{Cendes2021,Matsumoto2021-II}. Linear polarization was reported for jetted-TDE {\it Swift} J1644+57 in \cite{Wiersema2012}.  The observation was taken in the $Ks-$band about 18 days after {\it Swift}'s Burst Alert Telescope  {\it Swift-BAT} trigger yielding $\Pi=7.5\pm3.5\%$ (2$\sigma$). OGLE16aaa has polarization $\Pi=1.81\pm0.42$\% in the $V-$band \cite{Higgins2019}. This is the event most similar to AT 2020mot. However, the time of the polarization observation with respect to the peak optical emission is not known for OGLE16aaa, which prevents us from a more detailed comparison. Jetted-TDE {\it Swift} J2058.4+0516 was observed three times, $\sim81$, $\sim164$, and $\sim167$ days after the {\it Swift-BAT} trigger \cite{Wiersema2020}. The first observation yielded an upper limit of $\Pi<5.2\%$, followed by a detection ($\Pi=8.1\pm$2.5\%) and another upper limit ($\Pi<12.88\%$). Although the detection is within 2$\sigma$ of the initial upper limit, it would imply a rise in $\Pi$ similar to AT 2020mot. Polarization observations of AT 2019dsg were taken in the $V$-band and yielded a detection at peak ($\Pi=9.6\pm2.6$\%), and a $\Pi<2.7$\% upper limit 34 days later \cite{Lee2020}. The detection is again within $3\sigma$ of the upper limit, but would imply a decline in $\Pi$. None of the above observations have been corrected for the host-galaxy depolarization. Using archival observations we estimate the host galaxy magnitude of {\it Swift} J2058.4+0516 in the $H-$band $25.99$mag \cite{Pasham2015}. At the redshift of that TDE (z=1.1) even a massive elliptical host would be contained within the aperture. However, our estimates suggest that the host galaxy is $\sim300$ fainter than the TDE, hence any correction to $\Pi$ is negligible. It is possible that the intrinsic $\Pi$ for the remaining TDEs was higher. However, the lack of archival observations and the unknown aperture sizes used for the $\Pi$ estimation prevent us from estimating a correction factor.

\end{document}